# One Documentation Does Not Fit All

Case study of TensorFlow Documentation

Sharuka Promodya Thirimanne, Elim Yoseph Lemango, Giulio Antoniol*, Maleknaz Nayebi

*York University, *École Polytechnique de Montre'al*

{sharukat, elim17}@yorku.ca, Giulio.Antoniol@polymtl.ca, mnayebi@yorku.ca

*Abstract*—Software documentation provides guidance on the proper use of tools or services. With the rapid growth of machine learning libraries, individuals from various fields are incorporating machine learning into their workflows through programming. However, many of these users lack software engineering experience, affecting the usability of the documentation. Traditionally, software developers have created documentation primarily for their peers, making it challenging for others to interpret and effectively use these resources. Moreover, no study has specifically focused on machine learning software documentation or on analyzing the backgrounds of developers who rely on such documentation, highlighting a critical gap in understanding how to make these resources more accessible. In this study, we examined customization trends in TensorFlow tutorials and compared these artifacts to analyze content and design differences. We also analyzed Stack Overflow questions related to TensorFlow documentation to understand the types of questions and the backgrounds of the developers asking them. Further, we developed two taxonomies based on the nature and triggers of the questions for machine learning software.

Our findings showed no significant differences in the content or the nature of the questions across different tutorials. Our results show that 24.9% of the questions concern errors and exceptions, while 64.3% relates to inadequate and non-generalizable examples in the documentation. Despite efforts to create customized documentation, our analysis indicates that current TensorFlow documentation does not effectively support its target users.

## I. INTRODUCTION

Traditionally, software developers create documentation for their code primarily for other developers working in the same field. However, the increasing use of software code in diverse areas like Machine Learning (ML) means that many users who rely on this code documentation may not necessarily be software developers by profession. Code documentation, spanning various artifacts such as comments, reviews, pull requests, release notes, commit messages, and API documents, clarifies code functionality and reflects changes over time in natural language [1]. Studies identified common issues in software documentation, including insufficient or outdated content, unclear information, and lack of examples [2].

Software applications are expanding their scope, used across diverse domains from life sciences and economics to fashion and manufacturing. Hence, many developers are rather familiar with the application domain, while their educational and experience background is quite diverse when it comes to programming (i.e., they may know numerous programming languages and have experience in various domains in which programming is applied) and, more broadly, software engineering. This diversity necessitates accessible and understandable documentation tailored to different user needs. Skilled professionals have been relied upon for a variety of software engineering tasks, and their competence, profile, and skill are crucial in successfully completing development tasks promptly [3]. We argue for documentation to be effective, it should faithfully represent the software while accommodating users with a wide range of skills. This includes users with limited knowledge as well as those who are highly proficient [4].

Machine learning developers, system integrators, and end users come from diverse backgrounds, creating a varied user profile. Poor documentation of machine learning components can hinder usefulness, slow down time to market, and delay adoption. Ambiguous, outdated, and misleading documentation forces users to seek other resources, such as Stack Overflow. Various studies highlight the increasing trend of machine learning and Python queries on Stack Overflow [5]. Motivated by these challenges and aligned with the efforts on Software Engineering for Machine Learning (SE4ML) [6], we argue that the future of software documentation is to further align with the cross-disciplinary skills of machine learning users and even further to adjust and personalize this documentation to the use of individual developers. The potential of LLMs and generative AI to revolutionize software documentation and development practices is immense [7]. Embracing these technologies means transitioning from viewing development tools as mere facilitators to treating them as intelligent partners in the software creation process. This shift promises not only to enhance productivity and innovation but also to redefine the boundaries of software development [8].

We conducted an exploratory study to evaluate the current documentation of machine learning libraries [9], focusing on TensorFlow. We first examine the challenges developers face when using TensorFlow documentation and to further evaluate the extent to which the current TensorFlow documentation matches users' skill level. Based on our findings, we argue that the substantial discrepancy between user proficiency and documentation complexity needs special attention. This gap can lead to frustration, which is inefficient and far from an optimal use of time. We also explore generative AI's potential in software documentation design and implementation, envisioning a more personalized future for developers.

Our study makes a significant contribution as the first empirical research identifying developers' challenges with machine learning documentation and their skill levels, using TensorFlow as a case study. We developed a comprehensive taxonomy analyzing machine learning documentation issues and the profiles of users who posted those issues on Stack Overflow. This study aids developers, software engineers, and researchers in automated software documentation (e.g., [10]) by highlighting challenges faced by machine learning users.

In Section II, we outline our case study design. Section III details the empirical data used. In Section IV, we synthesized a taxonomy of identified problems and causes and extended our analysis to explore disparities between TensorFlow beginner and advanced documentation. Section V presents the study findings. Potential threats to the validity are addressed in Section VI. Related work is discussed in Section VII.

## II. STUDY DESIGN

Here, we discuss the context and protocols for our study.

### A. Case study Context Selection

Software documentation has been widely studied, primarily addressing developer challenges in the software engineering domain [11]. Our study positions itself at the intersection of software engineering and machine learning, focusing on Python and TensorFlow. Python is the preferred language for machine learning [12], making it the natural choice for our study. We selected TensorFlow due to two reasons:

**First,** We manually examined the documentation of popular ML libraries `PyTorch`, `Scikit-learn`, `TensorFlow`, `Keras`, `MXNet`, `Caffe`, `Theano`, and `CNTK`. Among them, TensorFlow is the only ML library that has tailored its documentation to accommodate users of varying skill levels. TensorFlow provides three documentation types, including Tutorials, API Documentation, and Community Translations. The tutorials section has two levels: beginner and advanced, with no detailed explanation of their differences. It considers the beginner tutorials the "best place to start with a user-friendly sequential API", while the advanced tutorials "provide a defined by-run interface for advanced research" [13]. Analysis of Stack Overflow tags for TensorFlow revealed Python as the most frequent tag, 1.9 times higher than Keras and 48.7% higher than Java.

**Second,** TensorFlow is a widely used machine learning library that was originally developed by `Google` and is now available as an open-source project. It provides wrappers in both Java and Python. Notably, TensorFlow is currently the only machine learning library adopting its documentation (even though partially and only for its tutorials) to the skill level of its users (beginner and advanced).

### B. Research Questions

In our study we answer three research questions. Through this case study, we investigate the below RQs;

**RQ1: How do the content of TensorFlow beginner and advanced tutorials compare?** *We focus on the TensorFlow tutorials that explicitly differentiate between beginner and advanced developers. We compare these in terms of the topics and subjects covered, the readability, word and sentence count of their content, and technicality (examples count and code snippets' size).*

**RQ2: How has the documentation for TensorFlow been discussed on Stack Overflow?** *We investigate whether users of TensorFlow face difficulties with the library's documentation and, if so, identify the nature and causes of these issues on Stack Overflow. We performed a systematic manual analysis to create taxonomies of the types of problems and the root causes of these issues.*

**RQ3: How do the years of experience and reputation of developers on Stack Overflow compare when asking questions about beginner vs advanced tutorials?** *Stack Overflow allows analysis of how developers of different skill levels ask about TensorFlow tutorials, helping assess whether separating them effectively serves the intended audiences. To assess if TensorFlow-defined tutorial levels align with developers' skills, we compared those who posted questions on Stack Overflow about either beginner or advanced tutorials using several metrics.*

## III. EMPIRICAL DATA

To answer **RQ1**, we gathered and used TensorFlow tutorials for both beginner and advanced users. Then we gathered questions about TensorFlow documentation from Stack Overflow to answer **RQ2**, and retrieved user profiles of those asking questions about TensorFlow tutorials to answer **RQ3**. Figure 1 depicts the process steps, which we reference below. The data gathering steps were annotated with yellow circles x⃝.

### A. TensorFlow Tutorials (**RQ1**)

TensorFlow provides tutorials for three types of artifacts: CORE, HUB, and ADD-ONS. Only the CORE product has tutorials for two competence levels: "beginner" and "advanced". Although TensorFlow officially maintains tutorials for HUB and ADD-ONS, they did not distinguish between different levels of competence. Therefore, for our analysis of **RQ1**, we focus specifically on the TensorFlow CORE product tutorials. To gather these tutorials in Step ①, we developed a Python scraper using the `Beautifulsoup` library to extract data from the TensorFlow website [14]. We then parsed the data to exclude the special characters (Step ②), page footers and headers, and ads and gathered the HTML tags for code snippets.

### B. Documentation Related Questions (**RQ2**)

Using StackExchange Data Explorer, we extracted Stack Overflow questions in two stages: (i) identifying keywords associated with documentation and (ii) retrieving and manually classifying documentation-related questions into taxonomies.

In Step ③, we queried questions with both the TensorFlow and Python tags and retrieved 45,866 questions. To focus on the documentation-related questions, we randomly sampled 1,000 questions and in Step ④, two researchers manually examined the questions posted on Stack Overflow and separated documentation-related questions from non-documentation ones. Among these 1,000 questions, we identified 124 questions as documentation-related. To increase the documentation-related questions count for our analysis, we extracted keywords frequently occurring in the questions about documentation. To do so, in Step ⑤ we pre-processed data by removing code blocks using HTML tags. We then removed standard `NLTK` stop words and numbers from the text. Then, we identified the Part-Of-Speech (POS) tags using `NLTK` and kept only nouns, adjectives, and verbs. The remaining words were then lemmatized, and calculated their TF-IDF values. We repeated these steps for the 876 non-documentation-related questions and ranked the words using their term frequency. We then excluded the frequently occurring words common between the two lists. To reverse the lemmatization, we used text surface realization with `pynlg`. As a result, in Step ⑥, we identified frequently used words in documentation-related questions: *documentation, document, doc, guide, tutorial, paper, instruction, official, lecture, book, website*.

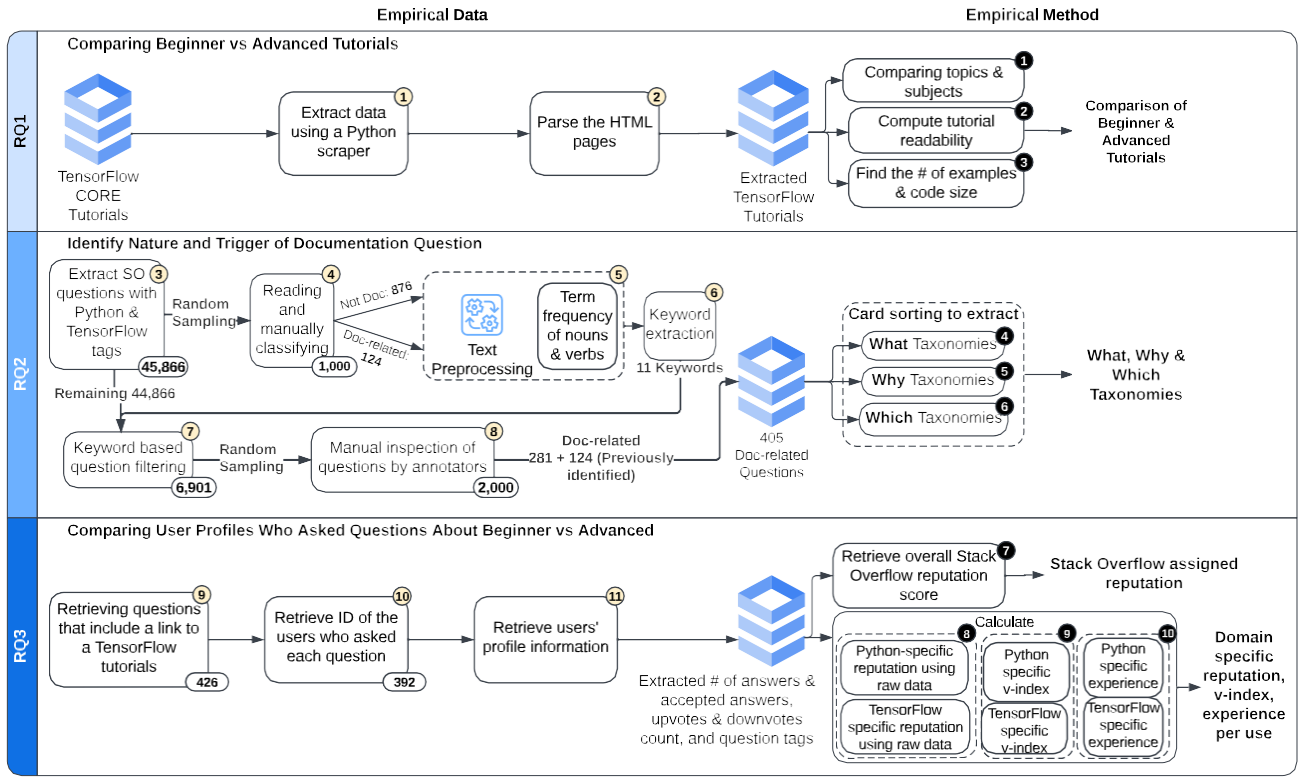

Fig. 1. Empirical Data collection (numbered in yellow) and research methodology (in black) for all the RQs

Then, we used these keywords to identify questions about documentation among the 44,866 Stack Overflow questions with Python and Tensorflow tags. Hence, in Step ⑦, we searched for questions with at least one of the documentation-related keywords above. We retrieved 6,901 questions that included at least one of these keywords in their text. Even if a question contains documentation-related keywords, it does not necessarily mean it pertains to documentation. To further verify this, we randomly sampled 2,000 questions from the 6,901 questions with documentation-related keywords. In Step ⑧, two annotators independently examined questions' relevancy to documentation. They identified 281 questions relevant to documentation. Having 124 documentation-related questions from our initial sampling in Step ④, we carried out our **RQ2** with an overall 405 documentation-related questions.

### C. Stack Overflow User Profiles (**RQ3**)

The Stack Exchange API provides endpoints to extract Stack Overflow user information. In Step ⑨, among those questions, we identified 426 questions that explicitly included a link (URL) to TensorFlow tutorial documentation. In Step ⑩, we retrieved the user profile of developers asking these questions. Out of 426 questions, we identified 392 unique user IDs (Only 33 users with more than one question). Finally, in Step ⑪ for each user we retrieved a range of data, including the post IDs, questions' creation date, link to the posts, number of answers, number of accepted answers, up-votes & down-votes count for the question, answers, and the owner ID.

## IV. EMPIRICAL METHOD

In this section, we explain our methodological steps. These steps are annotated with ⓧ in Figure 1.

### A. Content of TensorFlow's Beginners vs. Advanced Documentation (**RQ1**)

We compared the readability score, total sentence and word count, examples count, and code size for both beginner and advanced tutorials created by the TensorFlow community.

**Topic coverage:** ❶ When gathering data and comparing the "Beginner" and "Advanced" TensorFlow tutorials (Section III-A & Table I), two independent authors first reviewed and categorized subtopics to identify overlaps. They then compared the content for similarities or differences. The annotators then discussed and resolved any disagreements.

**Text readability:** It is proved that academic and technical writing that targets expert users is inherently more complex in reading compared to the general texts in magazines or Wikipedia targeting non-experts [15]. Hence, to deepen our analysis of TensorFlow tutorials, we evaluated if beginners' tutorials are more readable compared to advanced ones. ❷ We used the Flesch Reading Ease score as a widely used metric for readability [16]. The Flesch score ranges from 0 to 100, where the higher score indicates a text is easier to read. Flesch Reading Ease assumes that short sentences and words with fewer syllables are easier to understand. Here, a [0-30] score indicates a hard-to-comprehend text, while plain English normally scores in [60-70]. We used `Textstat Python` package to calculate Flesch Reading Ease for each tutorial.

**Number and size of code examples:** Different tutorials have a different level of technicality based on the topic they cover [17]. The technical example count could indicate the level of technicality in code documentation [18]. ❸ We counted the number of examples in a tutorial based on the descriptions right above each code block using the keyword "example". Then, one author manually reviewed 145 examples and verified their counts in TensorFlow tutorials. We further retrieved the size of code snippets in the tutorial by counting the lines of code provided in between the HTML `<code>` tags.

*B. Nature and Trigger of Documentation Questions (RQ2)*

We explained in Section III-B that as a result of multiple manual and automated steps, we identified 405 documentation-related questions. To answer **RQ2**, two annotators independently and manually analyzed the nature of the problems that users faced regarding TensorFlow documentation and their underlying causes. In each case, we defined what documentation the user is questioning. In our annotations, we answered:

**What** types of problems do developers encounter when using TensorFlow documentation?
**Why** did the issues occur when users were working with the TensorFlow library?
**Which** type of documentation was being discussed or referenced in the question?

Two annotators separately answered each of the above questions. Both annotators have over a year of industry software engineering experience and M.Sc. degrees in Software Engineering. To answer the "What" question and explore the nature of the problem, we used the defined categories by Islam et al. [19]. These categories defined 27 different types of questions commonly asked by Stack Overflow users about machine learning. ❹ Each developer independently conducted closed card sorting [20] to assign each question to one of the predefined categories to identify the nature of the problem and address the "What" question. The annotation process achieved a kappa agreement of 0.77 between the two annotators, indicating substantial agreement. Any conflicts between annotations were resolved by a moderator, who reviewed the mismatched annotations and finalized the categorizations, ensuring consistency within each category.

❺ To determine the cause of the problems and respond to the "why" question, we analyzed the entire question thread, including both the original question and any responses manually. In this case, two annotators independently performed an open card sorting [20] where each analyzed the question threads and grouped them based on the similarity in the causes of the problem. Then, they named each group with a representative title and created new categories as needed. After discussing, they selected the most fitting category names and resolved any disagreements, with the moderator finalizing the decisions.

❻ We also identified the documentation that was subject to question to answer "which" question and determine whether the user referred to official TensorFlow documentation (such as tutorials or API documentation) or third-party documentation. Although we did not have a predefined set of labels for these categories, they were factual decisions not subject to annotator bias. Therefore, the two annotators collaborated to define the "Which" categories without overlapping each other's work.

*C. Comparing Developers Posing Questions about Beginner and Advanced Tutorials (RQ3)*

We analyzed the profiles of users asking questions about TensorFlow tutorials to understand the relationship between TensorFlow-defined skill levels and developers' experience and knowledge. This analysis was done separately for developers' software engineering and machine learning backgrounds. A proficient programmer may be unfamiliar with the concepts of neural networks or machine learning. Conversely, an individual with solid technical and theoretical knowledge of machine learning and modeling is not necessarily an expert programmer. In this context, the questions they pose about software documentation could differ. We measured and compared developers based on three metrics in each field of Python programming and Machine learning using TensorFlow:

**Months of experience in the field** ❿ To determine the user's domain-specific experience, we computed the duration in months from their first Python or TensorFlow question posted date to the current date. This time frame represents their experience in the field in months.

**Reputation overall and in the field:** Stack Overflow assigns each user ❼ an overall reputation score based on their activities and the community's perception of their quality and importance. We compared these scores among the two groups. ❽ We used the formula defined by Stack Overflow with five attributes to compute in the field reputation [21], allowing users either increase or decrease their reputation. Activities that boost reputation, like upvoted questions (+5), upvoted answers (+10), or accepted answers (+15), increase reputation points, while downvoted questions (-2) or answers (-2) result in deductions. We also compared the overall reputations of users who asked questions about beginner or advanced tutorials. And, we computed the reputation using the same method, focusing on questions tagged as Python or TensorFlow.

**V-index in the field:** ❾ Wang et al. [21] introduced *v-index* as a metric better reflecting user competence instead of the reputation score. *v-index*, which is inspired by the academic h-index score measures the $v$ *number of answers* a user has posted, each with at least $v$ *number of upvote counts*. We calculated the v-index for each developer separately based on their questions with Python or TensorFlow tags. Finally, we used the Mann-Whitney test [22] to assess significant differences between users asking about beginner and advanced tutorials across these metrics.

## V. CASE STUDY RESULTS

Here, we present the findings of **RQ1**, **RQ2**, and **RQ3**.

### A. Tensorflow beginner vs. advanced tutorials (RQ1)

To distinguish between beginner and advanced TensorFlow tutorials, we first compared their titles and topics. We then examined the readability of tutorials, as technical documents are often harder to read, and evaluated the number of examples and lines of code in each tutorial.

*1) Comparison of titles and topics:* TensorFlow offers two beginner and 11 advanced tutorials. We gathered beginner and advanced tutorials of TensorFlow following the process explained in Section III-A. As a result, we identified 70 different topics across 13 tutorial titles. Table I lists the topics and their defined levels as published by TensorFlow. We manually compared these tutorials' titles, subtitles, and content with two independent researchers.

Table I has two primary topics within beginner-level tutorials: "ML basics with Keras" and "Load and preprocess data." Additionally, "Distributed training" is one of the 11 topics covered in the advanced-level tutorials. Notably, "ML basics with Keras" (beginner) and "Distributed training" (advanced) share a common sub-topic, under the **"Save and load"**. The beginner's "Save and load" tutorial covers various methods for saving TensorFlow models during and after training, including checkpoints, entire models, and loading saved models. In contrast, the advanced tutorial for "Save and load," found under "Distributed training", focuses on the specific logistics related to saving and loading models within distributed training models using distribution strategies. Our scanning through the content of the beginner and advanced tutorials for this subtopic demonstrated extensive differences in the subject matter. Table I shows the overlapping subtopic in teletype font.

*2) Analyzing tutorial's readability:* To assess the nature of TensorFlow tutorial documentation, we determined each tutorial's readability, sentence, and word count at beginner and advanced levels. Then, we statistically compared these tutorials using the Mann-Whitney test and Cohen's effect size. We selected the Mann-Whitney test [22] because it is a non-parametric method used to compare the means of two independent groups.

We summarized these findings in Table II. Also, Figure 2 displays the boxplot distribution of the Flesch readability score, the total number of sentences, words, the example count, and code size for each beginner and advanced tutorial. When comparing the group means using the Mann-Whitney tests, we identified a statistically significant difference in the Flesch readability ease between beginner and advanced tutorials ($p - value$ = 0.002). Figure 2 shows that the group means of readability score is significantly higher for beginner tutorials when compared with the advanced ones. In other words, the beginner tutorials are significantly easier to read.

Notably, no statistically significant differences were found in the total number of sentences ($p$ = 0.823) or words ($p$ = 0.616). This is reaffirmed by the visual representation in Figure 2-(b) & (c), which shows an overlap between the two notches of boxplots in each group. Notably, both the total sentence and word counts show a small effect size. Moreover, sentence count and word count show a positive effect size.

*3) Comparing example count and integrated code size:* We compared the beginner and advanced tutorials, using the lines of code and example count. Based on the results in Table II, there is a statistically significant difference in the total number of examples between beginner and advanced tutorials with a negative medium-size effect. However, there is no statistically significant difference in lines of code between beginner and advanced tutorials, and it has a small effect size. The more detailed analysis of the obtained results shows that even though the advanced tutorials have fewer examples, those examples contain a significantly higher number of lines of code.

> Analysis shows no substantial overlap in topics between documentation levels, and there are no significant differences in the total number of sentences, words, or lines of code. However, beginner tutorials include more examples. Readability scores vary, with beginners at 58.21 and advanced users at 54.28 on the Flesch Reading Ease scale.

### B. Questions about TensorFlow Documentation (RQ2)

We gathered a set of 405 TensorFlow documentation-related questions from Stack Overflow (Section III-B) and took a systematic approach (Section IV) to answer **RQ2**.

The below example[1] shows a user question linking directly to a TensorFlow documentation. The question is with regards to *tf.data: Build TensorFlow input pipelines* tutorial. The developer needs to switch between iterators of different shapes, but the TensorFlow data input pipelines guide [23] only covers switching between identical output shapes:

> TensorFlow documentation has examples regarding switching between datasets using `reinitialize_iterator` or `feedable_iterator`, but they all switch between iterators of the same output shape, which is not the case here. How to switch between training & validation sets using `tf.Dataset` & `tf.data.Iterator` in my case?

---
[1]https://stackoverflow.com/questions/51997426

TABLE I
TENSORFLOW TUTORIALS TOPICS, SUBJECTS, AND THE COMPETENCE LEVEL DEFINED BY GOOGLE.

| Title : Beginner | Topic | Count |
|---|---|---|
| ML basics with Keras | Basic image classification, Basic text classification, Text classification with TF-Hub, Regression, Overfit & underfit, `Save and load`, Tune hyperparameters with Keras Tuner | 7 |
| Load & preprocess data | Images, Videos, CSV, NumPy, pandas.Dataframe, TFRecord & tf.Example, Added formats with tf.io, Text, Unicode, Subword tokenization | 1 |
| **Title : Advanced** | **Topic** | **Count** |
| Customization | Tensors and operations, Custom layers, Custom training: walkthrough | 3 |
| Distributed training | Distributed training with Keras, Distributed training with DTensors, Using DTensors with Keras, Custom training loops, Multi-worker training with Keras, Multi-worker training with CTL, Parameter server training, `Save and load`, Distributed input | 9 |
| Vision | Computer vision, KerasCV, CNN, Image classification, Transfer learning & fine-tuning, Transfer learning with tf-hub, Data augmentation, Image segmentation, Object detection with tf-hub, Video classification, Transfer learning | 11 |
| Text | Text and natural language processing, Get started with Keras NLP, Text and NLP guide | 3 |
| Audio | Simple audio recognition, Transfer learning for audio recognition, Generate music with RNN | 3 |
| Structured data | Classify structured data with preprocessing layers, Classification on imbalanced data, Time series forecasting, Decision forest models, Recommenders | 5 |
| Generativ | Stable diffusion, Neural style transfer, DeepDream, DCGAN, Pix2Pix, CycleGAN, Adversarial FGSM, Intro to autoencoders, Variational autoencoder, Lossy data compression | 1 |
| Model optimization | Scalable model compression with EPR, TensorFlow model optimization | 2 |
| Model understanding | Integrated gradients, Uncertainty quantification with SNGP, Probabilistic regression | |
| Reinforcement learning | Actor-critic method, TensorFlow agents | 2 |
| TF.estimator | Premade estimator, Linear model, Keras model to estimator, Multi-worker training with estimator, Feature columns | 5 |

TABLE II
P-VALUE AND EFFECT SIZE RESULTS FOR READABILITY, TOTAL SENTENCES, AND WORDS, NUMBER OF EXAMPLES AND LINES OF CODE

| Figure | Attribute | p-Value | Effect Size |
|---|---|---|---|
| 3.(a) | Flesch Ease readability | *0.002** | - 0.676 |
| 3.(b) | Total # of sentences | 0.823 | + 0.165 |
| 3.(c) | Total# of words | 0.616 | + 0.073 |
| 3.(d) | Number of examples | *0.037** | - 0.730 |
| 3.(e) | Lines of code (LOC) | 0.400 | + 0.303 |

The annotation process revealed the following categories in the example, as discussed in Section IV-B:

**What?** Describes the issue - *data adaptation, which involves converting raw data into the library's required format.*
**Why?** Explains the reason for the issue - insufficient information on exceptions and alternatives when *users struggling with instructions in different domains.*
**Which?** Refers to the documentation - the official TensorFlow Guide for *CORE TensorFlow modules.*

In Table III, we summarized the categories with relevant Stack Overflow question examples.

*1) What? - Nature of the Questions:* We used the categories for machine learning queries on Stack Overflow, as originally proposed by Islam et al. [19]. The Kappa agreement between the two annotators was 0.77, considered a substantial agreement [24]. *Errors and exceptions* (24.9%), *model creation* (9.1%), and *parameter selection* (8.6%) are most frequently questioned areas about machine learning documentation. Analysis yielded just one question for each of the *data cleaning*, *model selection*, *robustness*, and *setup* categories.

It is understandable that many turn to Stack Overflow for solving *errors or exceptions*. The predominance of questions related to *model creation* and *parameter selection* among machine learning queries on Stack Overflow can be attributed to the inherently complex and iterative nature of these processes.

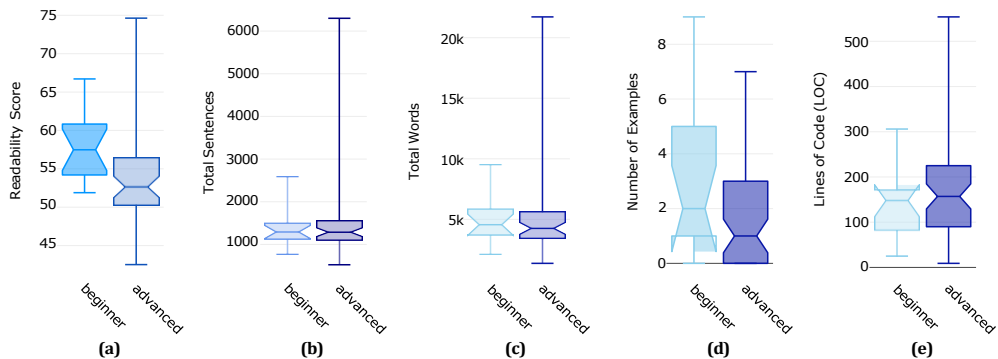

Fig. 2. Distribution of (a) readability score (b) sentence count (c) word count (d) examples count (e) lines of code with respect to the two tutorial levels (beginner and advanced)

TABLE III
TAXONOMY OF THE NATURE (WHAT?), REASONS (WHY?) OF QUESTIONS ABOUT TENSORFLOW DOCUMENTATION ON STACK OVERFLOW AND THE
DOCUMENTATION THAT WAS QUESTIONED (WHICH?).

| Taxonomy | Description | Cards% |
|---|---|---|
| **What?** | | |
| Error/Exception | All the errors related to Shape or Type Mismatch fall under this category. | 24.9% |
| Model creation | Is concerned with creation of new models | 9.13% |
| Parameter selection | Choosing appropriate values for the model to perform in the best possible manner | 8.64% |
| Data adaption | Obtaining the input data, reading it, and encoding it according to the specifications | 7.16% |
| Featuring | Reducing the data dimension to a focused feature space for useful information | 7.16% |
| Method selection | Is concerned with problems involved with the usage of APIs for performing validation | 6.91% |
| Prediction accuracy | Is concerned with issues related to model overfitting or underfitting | 5.67% |
| Performance | Is concerned with issues such as long training time or high memory consumption | 5.43% |
| Model load/store | Is concerned with loading the model on a disk and storing them | 4.44% |
| Model visualization | Is concerned with the visualization representation of images | 4.19% |
| Loss function | Issues on calculating the distance between actual and expected output, such as log loss. | 3.45% |
| Custom (non-ml) | Is the category defined to reference the outlier categories beyond the scope of ML | 3.45% |
| Optimizer | The function that helps with adjusting the weights and the learning rate of the model | 2.71% |
| Model conversion | Issues on transitioning between models trained with different libraries. | 1.72% |
| Output interpretation | Is the analysis of the results of the model in order to draw meaningful conclusions | 1.23% |
| Shape mismatch | Occurs when the matrix or tensor shape doesn't match the next neural network layer. | 0.49% |
| Model selection | Is concerned with choosing the best-fit model and API version | 0.24% |
| Data cleaning | Handling data values like null or missing values that could potentially cause problems | 0.24% |
| Robustness | Is concerned with the model's robustness to input dataset changes in reference to noise | 0.24% |
| Setup | All the conditions for the experiment, including the dataset, model, and parameters | 0.24% |
| **Why?** | | |
| Inadequate examples | Inadequate information on exceptions and alternatives within examples | 64.30% |
| Documentation ambiguity | Unclear or vague information in a document | 28.00% |
| Documentation completeness | Insufficient documentation on particular settings or configurations | 4.19% |
| Not ML Documentation | Requesting documentation irrelevant to machine learning tool or library | 1.48% |
| Lack of alternative solutions | Requiring a different solution to a problem | 0.74% |
| Document & code discrepancies | The difference between documentation provided and code implementation or syntax | 0.50% |
| Documentation replicability | Problems with replicating the results or examples provided in documentation | 0.25% |
| Request (additional) documentation | Lack of access to documentation due to deprecated links | 0.25% |
| **Which?** | | |
| TensorFlow tutorials | The official tutorials officially hosted on TensorFlow domain | 66.92% |
| Third-party documentation | Non-TensorFlow owned documentation (e.g., articles, blogs, reports, and websites). | 11.28% |
| TensorFlow GitHub documentation | TensorFlow Documentation provided on the project's open source GitHub repository | 6.82% |
| TensorFlow guides | TensorFlow hosted documentation are Jupyter notebooks run in Google Colab. | 6.30% |
| Scientific publications | Written works that report the results of scientific research | 2.09% |
| Multimedia tutorials | Including video tutorials on YouTube or online teaching platforms like Coursera | 0.52% |

*Model creation* involves designing and developing algorithms that learn from data, a task that requires a deep understanding of ML techniques, data structures, and extensive experimentation to optimize performance. Similarly, *parameter selection*, which involves tuning hyperparameters to improve model accuracy and efficiency, is a critical yet challenging step in the machine learning pipeline. Hyperparameters significantly impact model performance, but finding the right combination requires trial and error and specialized knowledge. Consequently, practitioners seek community guidance, increasing related questions on platforms like Stack Overflow.

*2) Why? - Motivation for the questions:* As a result of this process, one annotator identified 12 overall categories, while the other identified seven categories. The two annotators discussed the categories, merged the five common categories identified, and rephrased the category names. With the help of a mediator, they discussed the remaining differences and identified three more categories, then re-evaluated the questions for the task. As a result, we found eight categories to answer the "why" question (See Section IV-B for protocol details).

Based on our findings, the machine learning documentation is primarily subject to questions when developers apply a documented process or example in different application domains, accounting for 64.3% of the cases. In the second place, 28.0% of the questions are related to *document ambiguity* in the TensorFlow libraries. *Document completeness and mistakes* also triggered 4.2% of the questions. *Documentation including bugs and inconsistencies* (0.5%), *searching for alternative documentations* (0.7%), *replicating and taking the steps explained in a documentation* (0.2%), and *requesting and accessing other documentations* (0.2%) were also among the identified triggers. We also categorized 1.5% of the questions as unrelated to documentation, as their root cause was programming issues, though users still referenced the documentation.

Figure 3 depicts the heatmap on the intersection of the what (questions' nature) and why (the trigger of the questions). Overall, the "errors and exceptions" caused by "inadequate examples" have the highest proportion of 16.3% of the issues related to TensorFlow documentation. Secondly, 7.1% of the questions are due to "errors and exceptions" occurring because of "documentation ambiguity". Additionally, 5.4% of questions on "modal creation" and 5.1% on "parameter selection" arose due to "inadequate examples".

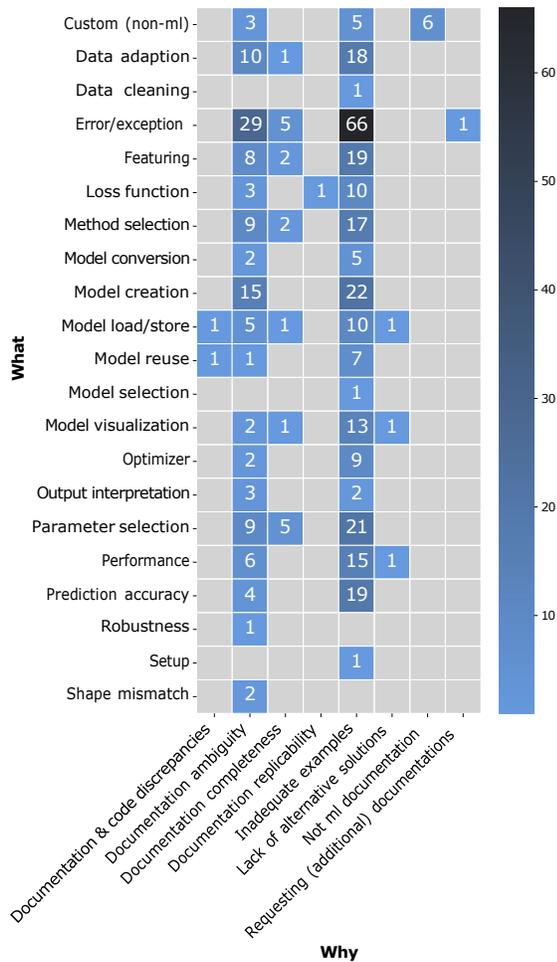

Fig. 3. Frequency of all questions on What area vs Why for Stack Overflow Questions about TensorFlow.

*3) Which? - The documentation subject to questions:* We also discussed the type of documentation used and referenced in the Stack Overflow questions. According to the data obtained incorporating an open card sorting (see Section IV-B), 97% of the questions we identified are related to TensorFlow documentation (API documentation, tutorial, or guide). Among them, the majority of the questions were regarding TensorFlow tutorials (66.9%). Moreover, the rest of the questions are subjected to TensorFlow guides (6.3%), TensorFlow GitHub documentation (6.8%), third-party documentation (11.3%), scientific publication (2.0%), and multimedia tutorials (0.5%). These observations revealed that the majority of the users face problems when using *official* TensorFlow tutorials or other official documentation.

TABLE IV
MANN WHITNEY TEST P-VALUE AND COHEN'S EFFECT SIZE

| Category | Attribute | p-Value | Effect Size |
|---|---|---|---|
| **Python** | 5.(b) Reputation | 0.060 | + 0.035 |
| | 5.(c) Experience in months | 0.036 | − 0.146 |
| | 5.(d) v-index | 0.198 | − 0.047 |
| **TensorFlow** | 5.(e) Reputation | 0.460 | + 0.107 |
| | 5.(f) Experience in months | 0.126 | − 0.010 |
| | 5.(g) v-index | 0.700 | + 0.064 |

> Nearly 25% of questions about TensorFlow documentation were about errors and exceptions due to insufficient or unclear examples. Also, majority (64.3%) of the questions were due to the inadequacy of examples and details about alternative solutions.

*C. Comparing Users Who Pose Questions about Beginner and Advanced Tutorials (RQ3)*

To address **RQ3**, we analyzed Stack Overflow user profiles to compare the background knowledge of users asking about beginner vs. advanced tutorials. Specifically, we examined overall reputation, v-index, reputation, and experience (in months) for Python and TensorFlow, conducting this analysis for users of each tutorial. Figure 4 shows the comparisons.

When comparing the Stack Overflow reputations assigned to users asking about beginner-level tutorials versus those inquiring about advanced-level tutorials, we found no significant difference. The analysis yielded a p-value of 0.121 and an effect size of 0.077. This finding remained insignificant when comparing their v-index in Python or TensorFlow. Using the Mann-Whitney test to compare the v-index between the two groups, we obtained a p-value of 0.198 with an effect size of −0.047 for Python, and a p-value of 0.7 with an effect size of 0.064 for TensorFlow, both of which were insignificant.

Further, we computed the Python and TensorFlow-specific reputations using the attributes listed in Section IV-C. This process was carried out separately for questions associated with Python or TensorFlow hashtags. However, we did not observe any significant difference in the reputation of users asking questions about beginner tutorials versus advanced tutorials in Python or TensorFlow. This is supported by the p-values of 0.06 and 0.460, and the relatively weak effect sizes of 0.035 and 0.107 for Python and TensorFlow, respectively. Conversely, when comparing the months of experience in Python and TensorFlow based on the time elapsed since each developer's first question with these tags, we found a significant p-value of 0.036 for Python. This indicates a notable difference in the distribution of months of experience in Python between users asking about beginner tutorials and those asking about advanced tutorials. Notably, developers asking beginner-level questions have more months of experience in Python. However, this p-value is accompanied by a relatively weak negative effect size (−0.146), leading us to consider this finding inconclusive. Furthermore, we obtained a p-value of 0.126 for months of experience in TensorFlow, indicating no significant difference, along with an effect size of −0.01. Table IV and Figure 4 summarize these comparisons.

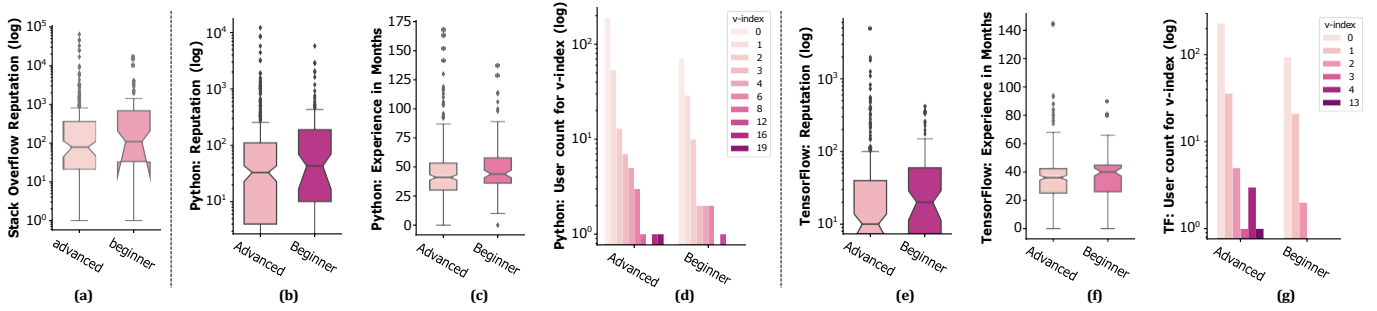

Fig. 4. (a) Users Stack Overflow reputation (b) Python-specific calculated reputation (c) Python-specific experience in months (d) Python-specific v-index (e) TensorFlow-specific calculated reputation (f) TensorFlow-specific experience in months and (g) TensorFlow-specific v-index.

> We observed no significant difference in developers' profiles when posing questions about beginner or advanced tutorials, considering overall reputation, Python or TensorFlow-specific reputation, or v-index

## VI. Threats to Validity of the Empirical Study

The metrics used could impose some *construct validity* on our results. We presented the Flesch Reading Ease [25] results. However, there are serveral metrics with slightly different algorithms to measure readability. We evaluated and compared the readability of beginner and advanced tutorials using three Gunning Fog Index, Flesch Reading Ease, and Flesch-Kincaid Grade metrics and found a high correlation between these values (min of 0.94 correlation and average of 0.957). Hence, we relied on the most popular metric, Flesch Reading Ease.

For *conclusion validity*, we analyzed reputation, v-index, and experience in **RQ3**, building on state-of-the-art methods [21]. However, user profiles may differ in other metrics not studied here. However, we believe analyzing state-of-the-art measures from Stack Overflow and distinguishing them by Python and TensorFlow skill sets mitigates this risk. Using months of experience as a metric may pose a threat to validity. Experienced users may create new accounts to ask advanced questions, and some may have switched domains, potentially misrepresenting their experience. However, the diverse data and user experience levels in our study help mitigate these effects. Moreover, we used the Mann-Whitney U test and our reliance on a single test for multiple comparisons raises the possibility of statistical errors. We mitigated this threat by adjusting the thresholds using Bonferroni correction.

For *external validity*, the case study focuses on Python as the software engineering domain. While switching to Java may alter observations, the impact is minimal since attributes are derived from Stack Overflow user profiles, a common feature across all domains. The formation of two taxonomies is prone to *external validity*. Establishing "fully correct" taxonomies is inherently challenging. We used established card sorting techniques to this end [20] and leveraged the strengths of both open and closed card sorting (**RQ2**). For *internal validity*, we hypothesized that a user's skills are independent; however, in reality, these skills are interdependent. This may impose some internal validity in our modeling while we believe the chances are low as this was not further investigated in the relevant studies on developers' experience analysis either [26].

## VII. Related Work

In the field of machine learning, documentation is frequently discussed, with an emphasis on reporting proper information alongside the training and testing data and code to reduce what is termed documentation debt [27]. Various efforts have aimed to improve machine learning documentation, primarily by better empirically explaining the data [27], [28]. Particularly, Chang et al. [29] surveyed 24 AI/ML practitioners, reflecting on the practices of implementing and operationalizing ML documentation in their workflows, identifying challenges, and addressing them. Conversely, there is a well-established body of research focused on software documentation. With the advent of machine learning, the software aspects of machine learning libraries have become the subject of various studies.

### A. Software Documentation and its challenges

Using Stack Overflow to identify developers' questions and pain points has been an established study design in software engineering [30], [31]. In particular, it was used to identify issues of software documentation [32], [33]. Aghajani et al. [32] performed an empirical study using mailing lists, Stack Overflow, and project repositories to identify issues developers faced using software documentation. They further performed a survey to evaluate the mined repository results against developers' perspectives [2]. Further, Treude et al. [34] suggested a framework to evaluate the quality of software documentation. Venigalla et al. [35] underscored documentation challenges by mining developers' sentiment in commit messages, which revealed that 45% expressed a "trust" emotion.

Software documentation, meeting developers' information needs passively, has prompted the development of automated tools and techniques for generating and enhancing documentation based on code and developers' activities [36], [37].

### B. Software Engineering for Machine Learning

MLOps, or Machine Learning Operations, is a set of practices that combines machine learning, DevOps (development and operations), and data engineering to automate and streamline the deployment and management of machine learning models in production [38]. Along with the MLOps, different levels and granularity of documentation have also been discussed. MLOps emphasize on documentation needed for each step of software life cycle [39]. Particularly the documentation of functional and non functional requirements for machine learning software has been discussed [40]. Warnett et al. [41] investigated the understandability of MLOps system architecture descriptions and found that semi-formal MLOps system diagrams improved comprehension. Furthermore, the iterative nature of ML application development, with its variety of artifacts, makes achieving comparability, traceability, and reproducibility challenging. Schlegel et al. [42] conducted a literature review on ML artifact management systems (AMS), identifying current AMSs and their functional and non-functional properties and the role of documentation.

This work builds upon the existing body of literature developed over the years [43]–[57], [57]–[79].

## VIII. CONCLUSION

Software documentation is traditionally written by developers for developers. However, its use by domain experts without software engineering backgrounds is growing. Along with the discussions and applications of *software engineering for machine learning*, we explored the match between software developers' skill levels and the documentation by conducting a case study on TensorFlow. We mined Stack Overflow questions related to these documentations, building a taxonomy of issues encountered by developers. We compared beginner and advanced tutorials, evaluating their content in terms of readability and technicality. Furthermore, we analyzed the profiles of users who posed questions on each tutorial. Our findings showed that TensorFlow clearly segregates beginner and advanced tutorials topics, with no advanced discussions on the same subjects. Interestingly, all tutorials attract questions from developers with varying skill levels. Notably, beginner tutorials receive extensive inquiries from well-seasoned developers. We argue that as machine learning software becomes more widely used across domains, its documentation should be tailored to accommodate a variety of skills.